\documentclass[12pt,aps,pre,preprint,showpacs,showkeys]{revtex4}

\usepackage{graphicx}
\usepackage{subfigure}
\usepackage{amsmath}
\usepackage{amssymb}
\usepackage{bm}

\begin{document}

\title{Magnetic quasi-long-range ordering in nematics due to competition between higher-order couplings}
\author{Milan \v{Z}ukovi\v{c}}
 \email{milan.zukovic@upjs.sk}
\author{Georgii Kalagov}
 \affiliation{Institute of Physics, Faculty of Science, P. J. \v{S}af\'arik University, Park Angelinum 9, 041 54 Ko\v{s}ice, Slovakia}
\date{\today}

\begin{abstract}
Critical properties of the two-dimensional $XY$ model involving solely nematic-like biquadratic and bicubic terms are investigated by spin-wave analysis and Monte Carlo simulation. It is found that, even though neither of the nematic-like terms alone can induce magnetic ordering, their coexistence and competition leads to an extended phase of magnetic quasi-long-range order phase, wedged between the two nematic-like phases induced by the respective couplings. Thus, except for the muticritical point, at which all the phases meet, for any finite value of the coupling parameters ratio there are two phase transition: one from the paramagnetic phase to one of the two nematic-like phases followed by another one at lower temperatures to the magnetic phase. The finite-size scaling analysis indicate that the phase transitions between the magnetic and nematic-like phases belong to the Ising and three-state Potts universality classes. Inside the competition-induced algebraic magnetic phase the spin-pair correlation function is found to decay even much more slowly than in the standard $XY$ model with purely magnetic interactions. Such a magnetic phase is characterized by an extremely low vortex-antivortex pair density attaining a minimum close to the point at which the biquadratic and bicubic couplings are of about equal strengths and thus competition is the fiercest.
\end{abstract}



\maketitle

\section{Introduction}

A standard two-dimensional $XY$ model with the Hamiltonian ${\mathcal H_1}=-J_1\sum_{\langle i,j \rangle}\cos(\phi_{i,j})$, where the interaction $J_1>0$ is limited to nearest-neighbor pairs forming the angle $\phi_{i,j}=\phi_{i}-\phi_{j}$, is well known to show a topological Berezinskii-Kosterlitz-Thouless (BKT) phase transition. The quasi-long-range-order (QLRO) BKT phase arises due to the vortex-antivortex pairs unbinding~\cite{bere71,kost73} and it is characterized by an algebraically decaying correlation function $g_1(r) = \langle \cos(\phi_{0}-\phi_{r}) \rangle \sim r^{-\eta_1}$. It's generalization to nematics can be obtained by replacing the magnetic spin-spin interaction by (pseudo)nematic higher-order terms described by the Hamiltonian
 ${\mathcal H_q}=-J_q\sum_{\langle i,j \rangle}\cos(q\phi_{i,j})$, where $q=2,3,\hdots$.
Due to the fact that the partition function of the latter can be mapped onto the former by the transformation $q\phi_{i} \to \phi_{i}$, also the nematics show a (nematic) QLRO phase with the correlation function $g_q(r) = \langle \cos q(\phi_{0}-\phi_{r}) \rangle  \sim r^{-\eta_q}$ and the same order-disorder transition temperature~\cite{carme87}. While the correlation function $g_1$ is related to the ferromagnetic ordering in which spins have a common direction, $g_q$ is related to the nematic term that does not induce directional but only axial alignments with angles $2k\pi/q$, where $k$ is an integer and $k \leq q$. Consequently, in the nematics there is no magnetic ordering and $g_1$ is expected to decay exponentially.  

Several models that combine the bilinear and higher-order terms have been proposed. Their motivation was either theoretical curiosity (critical properties and universality) or various experimental realizations (e.g., liquid crystals~\cite{lee85,geng09}, superfluid A phase of $^3{\rm He}$~\cite{kors85}, or high-temperature cuprate superconductors~\cite{hlub08}, DNA packing~\cite{grason08}, quasicondensation in atom-molecule, bosonic mixtures~\cite{bonnes12,bhas12,forg16}, and structural phases of cyanide polymers~\cite{cairns16,zuko16}. The most studied model, that included the bilinear and biquadratic terms, i.e., the system with the Hamiltonian ${\mathcal H}={\mathcal H_1}+{\mathcal H_2}$, has been shown~\cite{lee85,kors85,carp89,shi11,hubs13,qi13} to lead to the separation of the magnetic phase at lower and the nematic phase at higher temperature, for sufficiently large biquadratic coupling. The high-temperature phase transition to the paramagnetic phase was determined to belong to the BKT universality class, while the magnetic-nematic phase transition had the Ising character. 

Further generalization of the nematic term, which leads to the Hamiltonian ${\mathcal H}={\mathcal H_1}+{\mathcal H_q}$, where $q>2$, surprisingly revealed a qualitatively different phase diagram if $q \geq 5$~\cite{pode11,cano14,cano16}. The newly discovered ordered phases appeared as a result of the competition between the ferromagnetic and pseudonematic couplings and the respective phase transitions were determined to belong to various (Potts, Ising, or BKT) universality classes.

There have been several other modifications and generalizations of the $XY$ model involving higher-order terms, such as taking the $k$-th order Legendre polynomials of the bilinear term (${\mathcal H}=-\sum_{\langle i,j \rangle}P_k(\cos(\phi_{i,j}))$)~\cite{fari05,berc05} or in another nonlinear form (${\mathcal H}=2J\sum_{\langle i,j \rangle}(1-[\cos^2(\phi_{i,j}/2)]^{p^2})$)~\cite{doma84,himb84,blot02,sinh10a,sinh10b} as well as the generalization by inclusion of up to an infinite number of higher-order pairwise interactions with an exponentially decreasing strength (${\mathcal H}=-\sum_{\langle i,j \rangle}\sum_{k=1}^{p}J_{k}\cos^{k}\phi_{i,j}$, where $J_k=\alpha^{-k}$ and $\alpha>1$)~\cite{zuko17}. The main focus was the possibility of the change of the BKT transition to first order, the existence of which in the former model was rigorously proved for sufficiently large values of the parameter $k$~\cite{ente02,ente05}.  

In the present study we consider the model that involves purely nematic (biquadratic and bicubic) terms and completely lacks the magnetic interaction (${\mathcal H}={\mathcal H_2}+{\mathcal H_3}$). In spite of the fact that neither of the nematic interactions alone can induce magnetic ordering, we demonstrate that their competition leads to an extended magnetic QLRO phase with the spin-pair correlation function $g_1(r) \sim r^{-\eta_{1}^{eff}}$ decaying even much more slowly than in the standard $XY$ model.

\section{Model and Methods}

The studied model Hamiltonian on a square lattice takes the following form
\begin{equation}
\label{Hamiltonian}
{\mathcal H}=-J_2\sum_{\langle i,j \rangle}\cos(2\phi_{i,j})-J_3\sum_{\langle i,j \rangle}\cos(3\phi_{i,j}),
\end{equation}
where $\phi_{i,j}=\phi_{i}-\phi_{j}$ is an angle between the nearest-neighbor spins and the respective exchange interactions are considered as follows: the biquadratic $J_2 \equiv J$ and the bicubic $J_3=1-J$, with $J \in [0,1]$.

\subsection{Spin wave approximation}
We are interested in a large-scale behavior of the pair correlation function $ g_q(x_1-x_2) \equiv \left\langle \cos q ( \phi(x_1) -  \phi(x_2) ) \right \rangle $, where $x$ is the coordinate vector of the $i$th spin, $q \in \mathbb{N}$ and brackets $\left\langle \dots  \right \rangle $ denote an average over possible spin configurations. The general form of the model Hamiltonian~(\ref{Hamiltonian}), involving higher-order terms up to the $n$th order, is given by 
\begin{equation}
{\mathcal H} = -\sum_{\left\langle i,j \right\rangle} \sum_{q=1}^{n} J_q \cos(q \, \phi_{i,j}).
 \end{equation}
Then passing to the continuous limit within the spin-wave approximation, we arrive at the effective Hamiltonian 
 \begin{equation}
{\mathcal H}^{eff} = \frac{J^{eff}}{2} \int \mathrm{d}^2 x [\nabla \phi(x)]^2,
 \end{equation}     
where the effective coupling $J^{eff} \equiv \sum_{q=1}^{n} J_q q^2$. Direct computation of the correlation function in the large scale region $|x_1 - x_2| \gg a$, where $a$ is the lattice vector, using the effective Hamiltonian gives
 \begin{equation}
g_q(x_1 - x_2) = \int \prod_{x} \mathrm{d} \phi(x) \exp\left( -{\mathcal H}^{eff} + i q [\phi(x_1)  - \phi(x_2)]\right)
 \end{equation}
and leads to the result 
 \begin{equation}
g_q(x_1 - x_2) = C_0 \exp\left( -\frac{ q^2}{2 \pi J^{eff}} \ln \frac{|x_1-x_2|}{a}\right)  \propto \left(\frac{a}{|x_1 - x_2|}  \right)^{\eta_{q}^{eff}}.     
 \end{equation}
 Here $C_0$ is an unessential constant and the critical exponent $\eta_{q}^{eff} = q^2T/(2 \pi J^{eff})$. Note, in order to perform the Gaussian integration correctly, the effective exchange interaction $J^{eff}$ has to be a positive quantity, i.e. $\sum_{q=1}^{n} J_q q^2 > 0$. 

\subsection{Monte Carlo}
Spin systems on a square lattice of a side length $L$ with the periodic boundary conditions are simulated by employing the Metropolis algorithm. We take $2 \times 10^5$ Monte Carlo sweeps (MCS) for thermal averaging after discarding another $4\times 10^4$ MCS to bring the system to the equilibrium. Temperature dependencies of various thermodynamic quantities are obtained by cooling the system from the temperature $T$ (measured in units $J/k_B$, where $k_B$ is the Boltzmann constant) in the paramagnetic phase down to lower temperatures with the step $\Delta T=0.025$, using the last configuration obtained at the previous temperature to initialize the simulation at the next temperature.  

In order to accurately estimate critical exponents between different phases and thus reliably determine the universality classes of the respective transitions, we also perform finite-size scaling (FSS) analysis by using the reweighting techniques~\cite{ferr88,ferr89} for the lattice sizes $L=24-144$. Since the integrated autocorrelation time is considerably enhanced close to the transition point (found to be of the order of $\propto 10^4$ MCS for the largest lattice size), we increase the number of MC sweeps to be used in the reweighting up to $10^7$ after discarding $2 \times 10^6$ MCS for thermalization. Statistical errors are evaluated using the $\Gamma$-method~\cite{wolf04}. 

The quantities of interest include the internal energy per spin $e=\langle {\mathcal H} \rangle/L^2$, 
the specific heat per spin $c$
\begin{equation}
c=\frac{\langle {\mathcal H}^{2} \rangle - \langle {\mathcal H} \rangle^{2}}{L^2T^{2}},
\label{c}
\end{equation}
the QLRO parameters $o_q$, $q=1,2,3$,
\begin{equation}
o_q=\langle O_{q} \rangle/L^2=\left\langle\Big|\sum_{j}\exp(iq\phi_j)\Big|\right\rangle/L^2,
\label{oq}
\end{equation}
and the corresponding susceptibilities $\chi_{o_q}$
\begin{equation}
\label{chi_oq}\chi_{o_q} = \frac{\langle O_{q}^{2} \rangle - \langle O_{q} \rangle^{2}}{L^2T},
\end{equation}
where $O_q$ represents the magnetization $M$, for $q=1$, and the nematic parameters $N_q$, for $q=2,3$. Furthermore, we can calculate the following quantities:
\begin{equation}
\label{D1}D_{1q} = \frac{\partial}{\partial \beta}\ln\langle O_{q} \rangle = \frac{\langle O_{q} {\mathcal H}
\rangle}{\langle O_{q} \rangle}- \langle {\mathcal H} \rangle,
\end{equation}
\begin{equation}
\label{D2}D_{2q} = \frac{\partial}{\partial \beta}\ln\langle O_{q}^{2} \rangle = \frac{\langle O_{q}^{2} {\mathcal H}
\rangle}{\langle O_{q}^{2} \rangle}- \langle {\mathcal H} \rangle.
\end{equation}

At second-order phase transitions the above quantities scale with the system size as
\begin{equation}
\label{fss_o}o_q(L) \propto L^{-\beta/\nu},
\end{equation}
\begin{equation}
\label{fss_chi}\chi_{q}(L) \propto L^{\gamma/\nu},
\end{equation}
\begin{equation}
\label{fss_D1}D_{1q}(L) \propto L^{1/\nu},
\end{equation}
\begin{equation}
\label{fss_D2}D_{2q}(L) \propto L^{1/\nu}.
\end{equation}

In the BKT phase the exponent $\eta$ of the algebraically decaying correlation function can be obtained from FSS of the parameter $o_q$, by using the hyperscaling relation $2\beta/\nu=\eta$, in the form
\begin{equation}
\label{m_FSS}
o_q(L) \propto L^{-\eta/2},
\end{equation}
or from FSS of the susceptibility, by using the hyperscaling relation $\eta=2-\gamma/\nu$, in the form
\begin{equation}
\label{xi_FSS}
\chi_q(L) \propto L^{2-\eta}.
\end{equation}
 
A proper order parameter for the algebraic BKT phase is the helicity modulus $\Upsilon$ (or spin wave stiffness)~\cite{fish73,nels77,minn03}, which quantifies the resistance of the systems to a twist in the boundary conditions. It is defined as the second derivative of the free energy density of the system with respect to the twist $\tau$ along one boundary axis, which, for example, for the present model with the Hamiltonian~(\ref{Hamiltonian}) results in the following expression
\begin{equation}
\label{helicity}
\Upsilon = \frac{1}{L^2}\left\langle \sum_{\langle i,j \rangle_x} 4J\cos (2\phi_{i,j}) + 9(1-J)\cos (3\phi_{i,j})\right\rangle - \frac{\beta}{L^2}\left\langle\Big[\sum_{\langle i,j \rangle_x} 2J\sin (2\phi_{i,j}) + 3(1-J)\sin (3\phi_{i,j})\Big]^2\right\rangle,
\end{equation}
where the summation $\sum_{\langle i,j \rangle_x}$ is taken over the nearest neighbors along the direction of the twist.

We also measure the presence of topological excitations directly from MC simulations. In particular, in each equilibrium configuration we detect all vortices and antivortices, as topological objects which correspond to the spin angle change by $2\pi$ and $-2\pi$, respectively, going around a closed contour enclosing the excitation core. Then, we calculate the vortex density $\rho$ by performing thermodynamic averaging and normalizing by the system volume.

\section{Results}
\begin{figure}[t!]
\centering
\includegraphics[scale=0.52,clip]{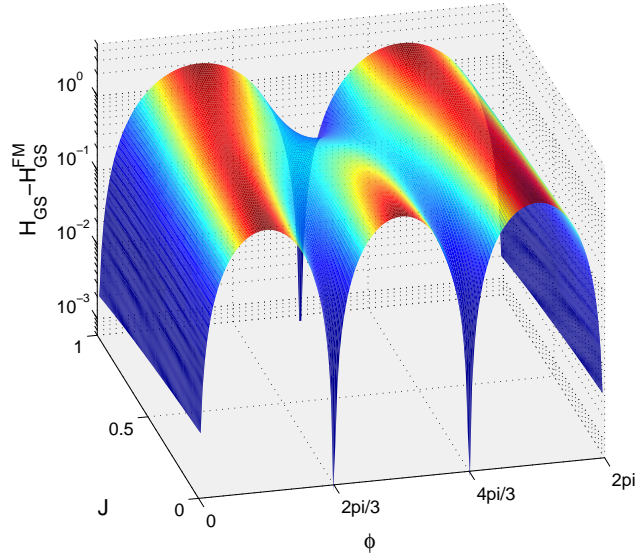}
\caption{(Color online) The difference of the ground-state energies ${\mathcal H_{GS}}-{\mathcal H_{GS}^{FM}}$, where ${\mathcal H_{GS}}=-J\cos(2\phi)-(1-J)\cos(3\phi)$ and ${\mathcal H_{GS}^{FM}}=-1$, shown as a function of the parameters $J$ and $\phi$.}\label{fig:gs}
\end{figure}

Let us first examine the ground-state behavior. In the limiting values of the coupling parameter $J$ the system shows nematic-like orderings with the adjacent spins having a
phase difference of $2k\pi=q$, where $k \leq q$ is an integer and $q=2$ ($q=3$) for $J=1$ ($J=0$)~\cite{cano14}. Nevertheless, within $0 < J < 1$ the energetically preferred arrangements becomes the ferromagnetic one. It is apparent from Fig.~\ref{fig:gs}, which shows the difference between noncollinear states energies, given by the functional ${\mathcal H_{GS}}=-J\cos(2\phi)-(1-J)\cos(3\phi)$, and the energy of the ferromagnetic state ${\mathcal H_{GS}^{FM}}$, in the $J-\phi$ plane.

\begin{figure}[t!]
\centering
\includegraphics[scale=0.5,clip]{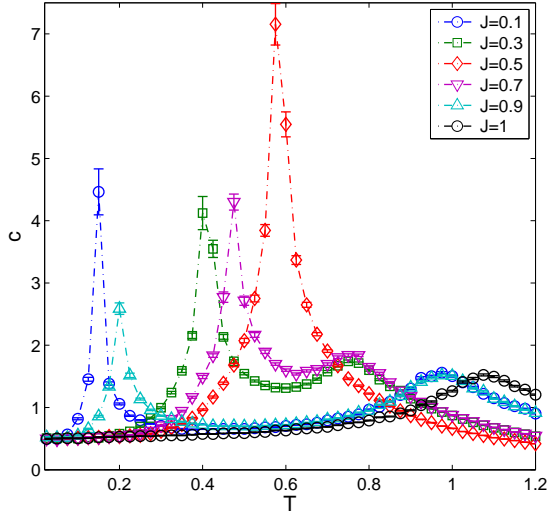}
\caption{(Color online) Temperature variations of the specific heat $c$, for different values of $J$ and $L=24$.}\label{fig:c-T}
\end{figure}

\begin{figure}[t]
\centering
    \subfigure{\includegraphics[scale=0.45]{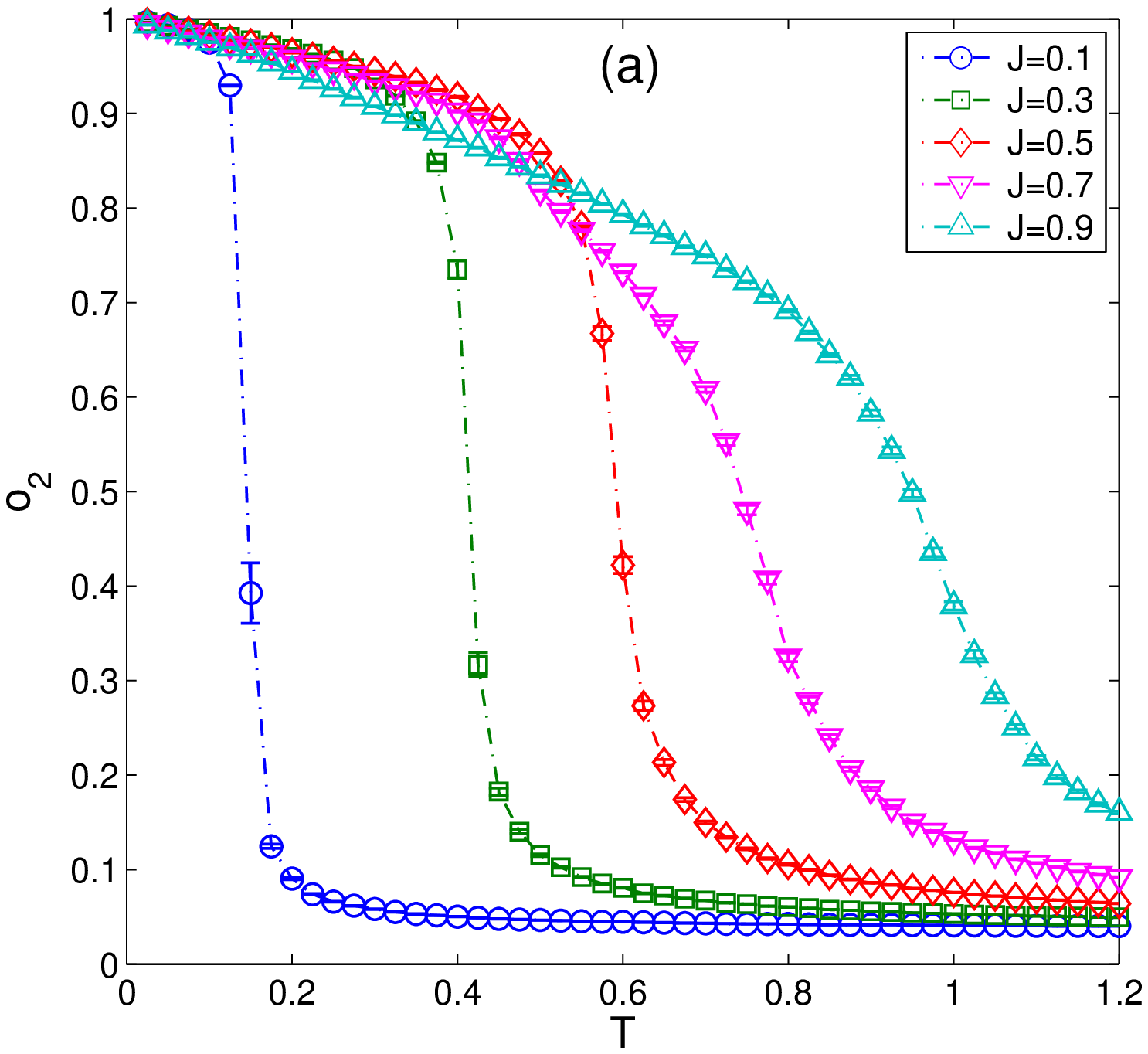}\label{fig:q2-T}}
		\subfigure{\includegraphics[scale=0.45]{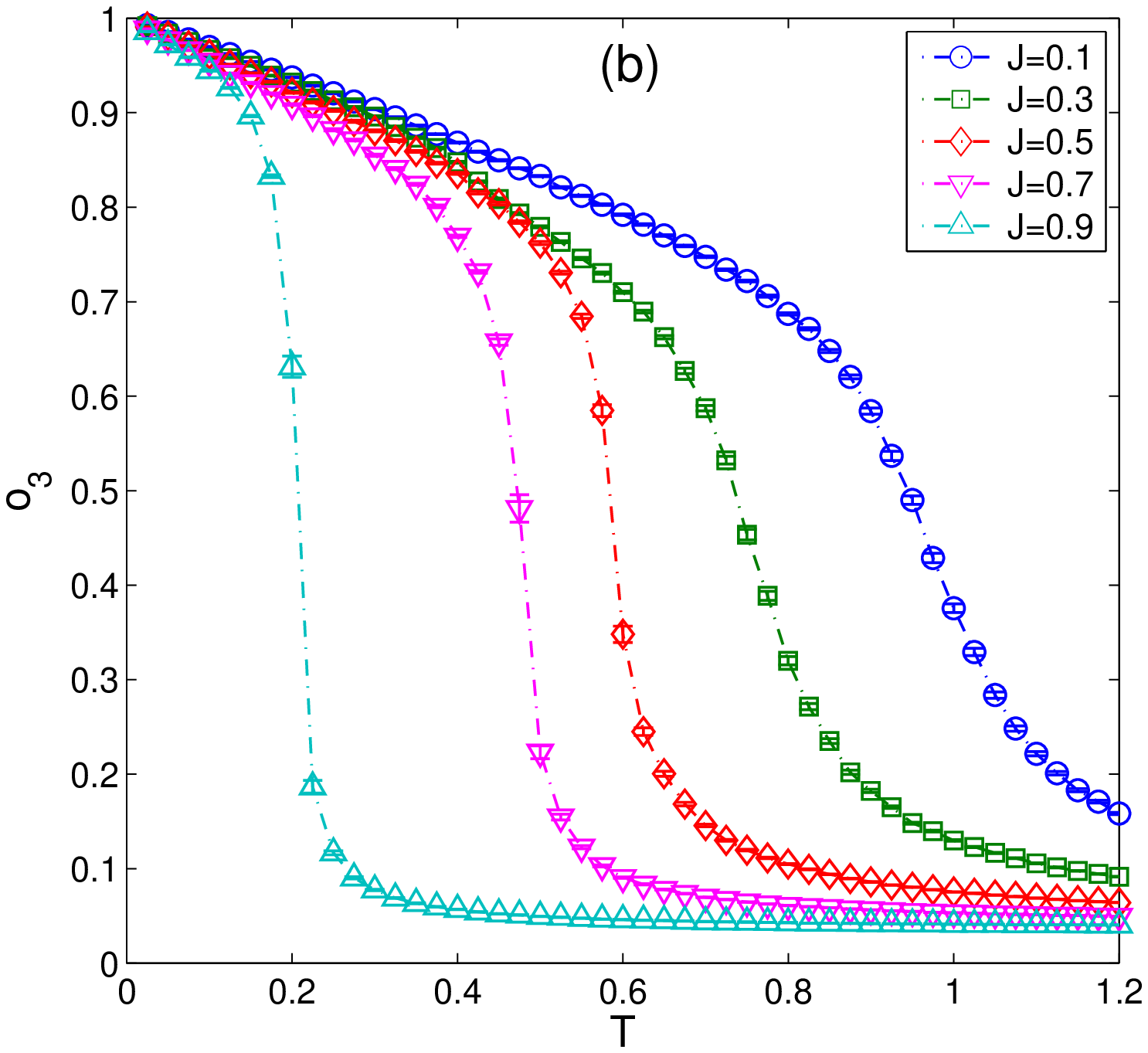}\label{fig:q3-T}}
		\subfigure{\includegraphics[scale=0.45]{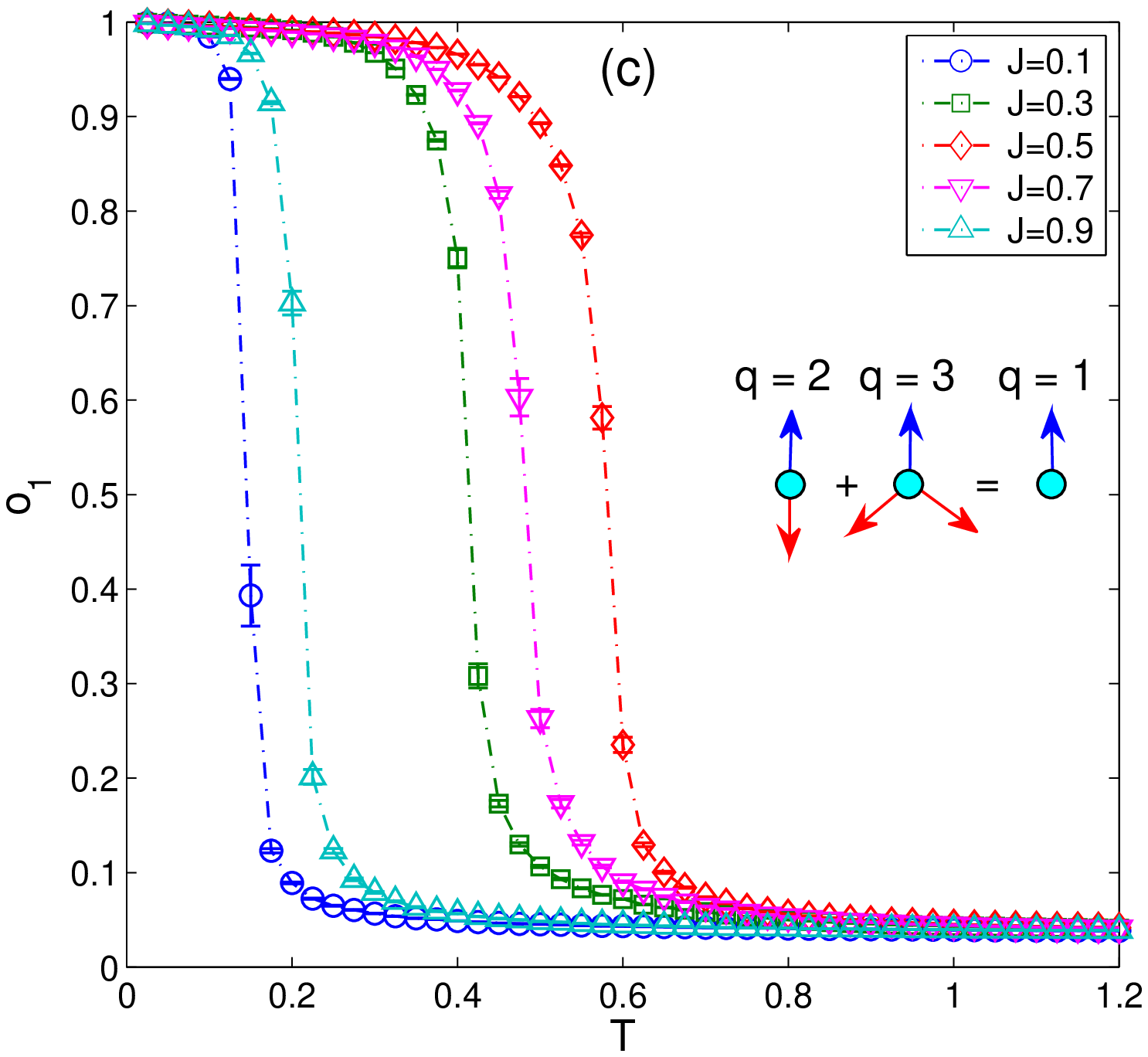}\label{fig:q1-T}}
		\subfigure{\includegraphics[scale=0.45]{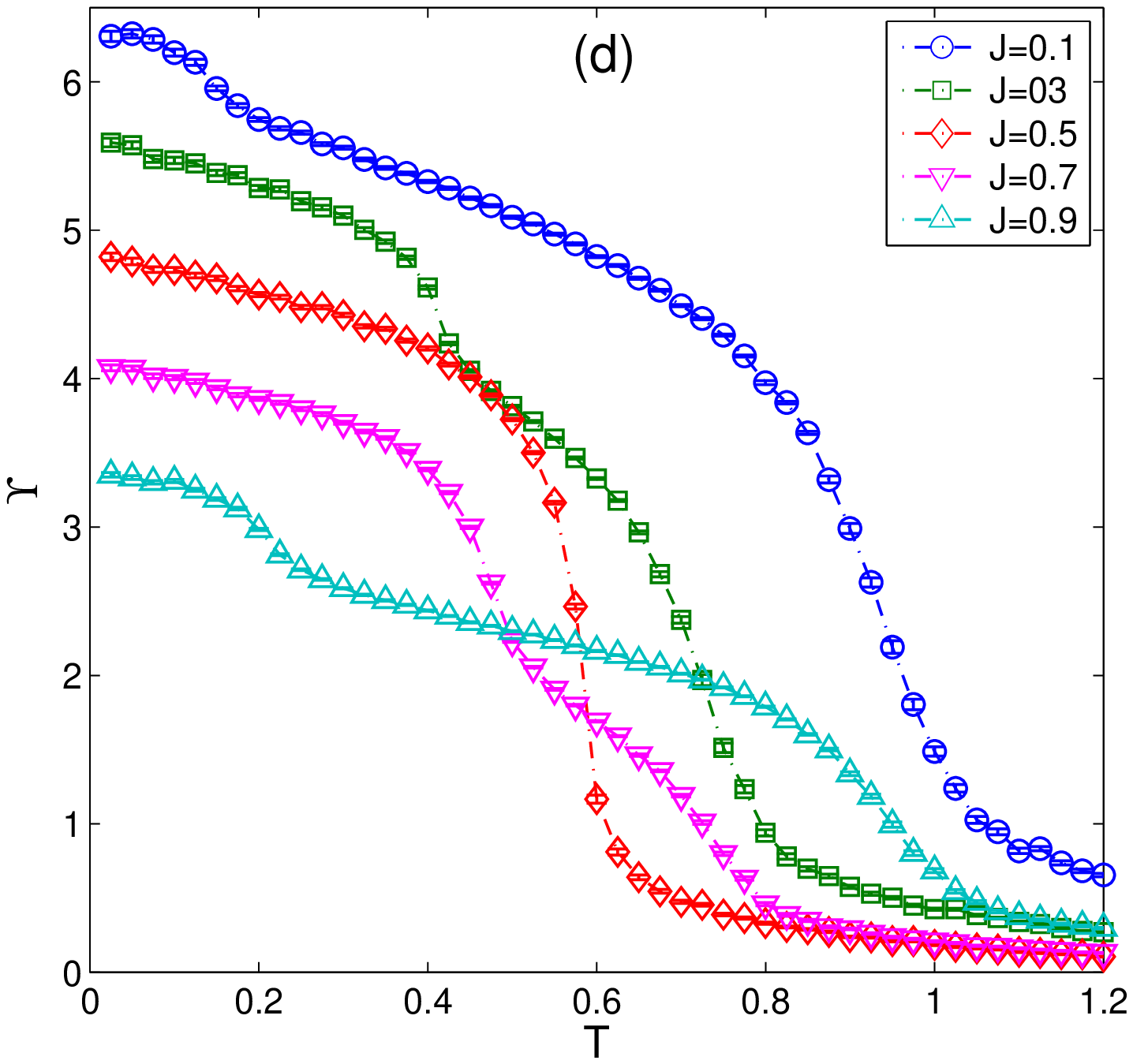}\label{fig:hm-T}}
\caption{(Color online) Temperature variations of the order parameters $q_2$, $q_3$, $q_1$, and $\Upsilon$, for different values of $J$ and $L=24$.}
\label{fig:x-T}
\end{figure}

In the following we show that for $0 < J < 1$ the ferromagnetic ordering also extends to finite temperatures with the crossover to the paramagnetic state either through one of the nematic-like states or directly. The double-peak structure of the specific heat behavior in Fig.~\ref{fig:c-T} indicates the presence of two phase transitions (except for $J=0.5$) and the character of the respective phases can be judged from the temperature dependencies of the respective order parameters for different values of $J$, presented in Fig.~\ref{fig:x-T}. The less prominent rounded high-temperature peaks in the specific heat curves are related to the order-disorder transitions between the respective nematic and the paramagnetic phases and are known to belong to the BKT universality class~\cite{pode11}. On the other hand, the low-temperature peaks look sharper and signify a different kind of the phase transitions that occur between the nematic and ferromagnetic phases. The latter arise due to the competition between the two kinds of the nematic interactions, as schematically illustrated in the inset of Fig.~\ref{fig:q1-T}. As thermal fluctuations are suppressed at sufficiently low temperatures both the collinear (equally allowing parallel and antiparallel states) and noncollinear arrangements are disfavored and all the spins align in the same direction, giving rise to the magnetic order parameter $o_1$. The resulting ferromagnetic phase extends to the highest temperature of $T \approx 0.6$, corresponding to $J \approx 0.5$. This is the point at which the effect of both nematic interactions is about equal, and at which the system appears to enter directly the paramagnetic phase. The successive phase transitions away from $J \approx 0.5$ are also reflected in two anomalies in the helicity modulus $\Upsilon$, shown in Fig.~\ref{fig:hm-T}.

\begin{figure}[t!]
\centering
		\subfigure{\includegraphics[scale=0.45,clip]{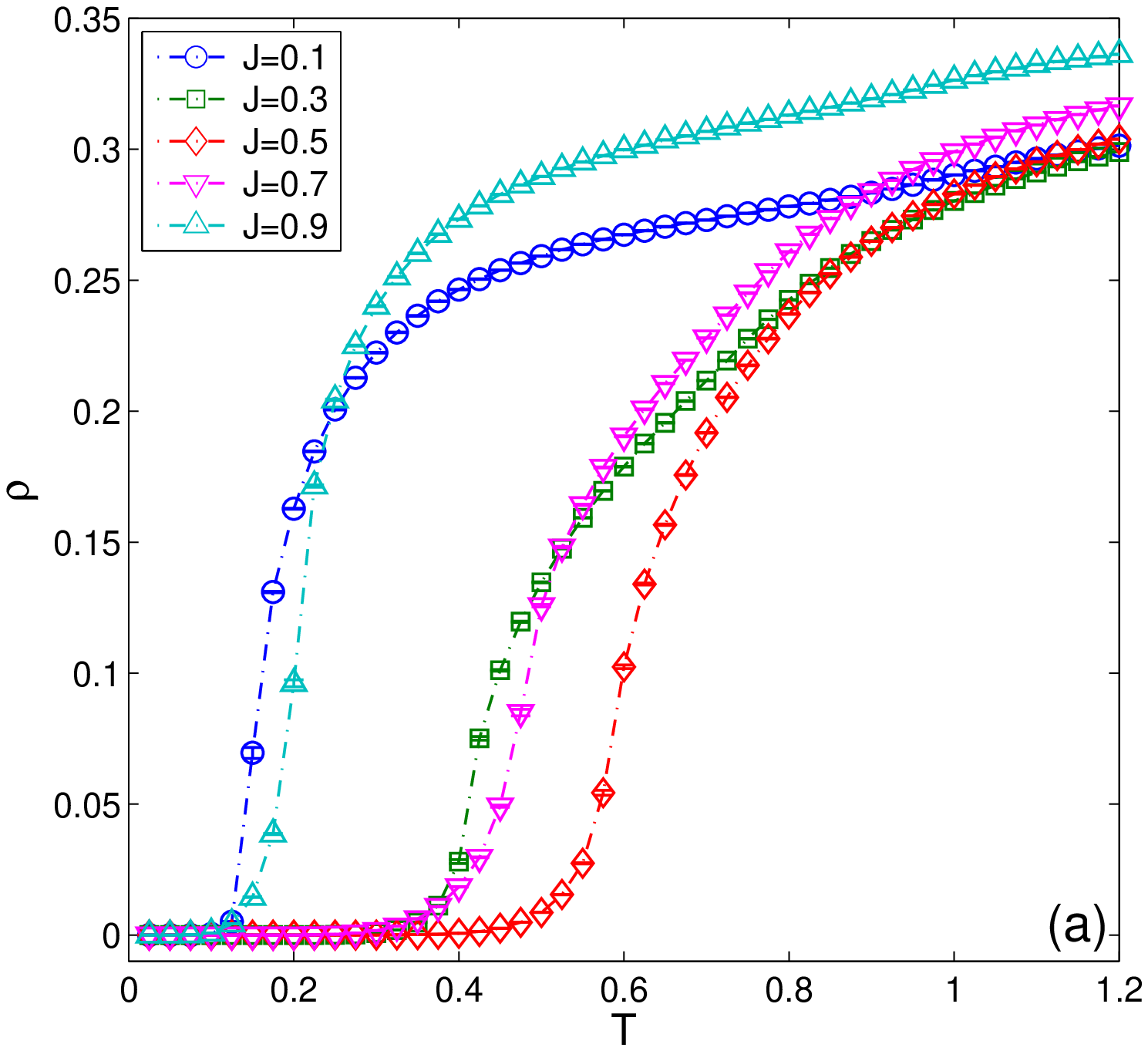}\label{fig:rho-T}}
    \subfigure{\includegraphics[scale=0.45,clip]{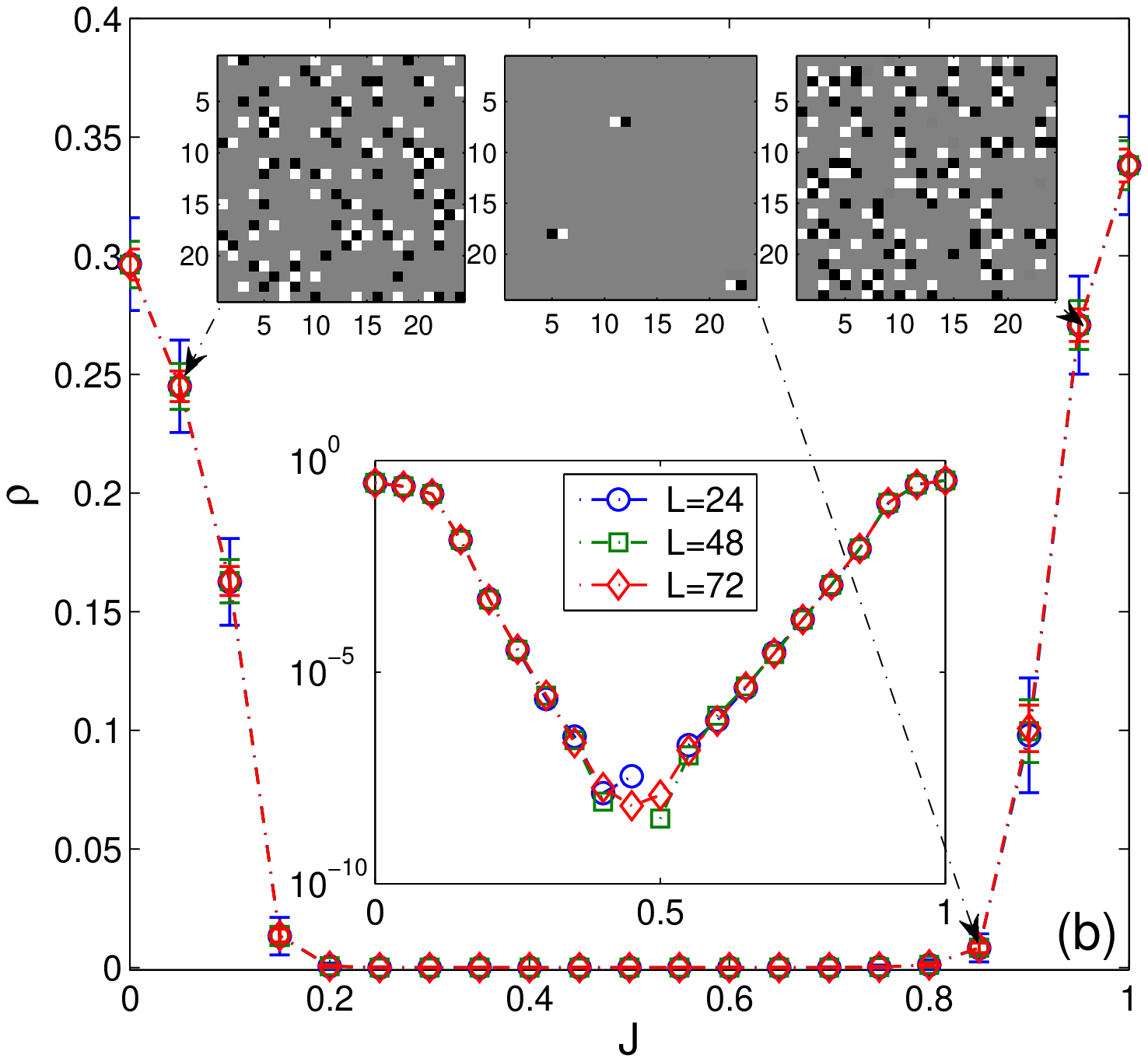}\label{fig:rho-J}}
\caption{(Color online) (a) Temperature variation of the vortex density $\rho$, for several values of $J$. (b) $\rho$ as a function of $J$ at a fixed temperature $T=0.2$, for three different values of the lattice size $L$. The insets show typical snapshots depicting vortices (white squares) and antivortices (black squares), outside ($J=0.05$ and $0.95$) and inside ($J=0.85$) the ferromagnetic phase.}\label{fig:nvrt}
\end{figure}

Fig.~\ref{fig:nvrt} shows the variation of the vortex density $\rho$ in the model parameter space. In Fig.~\ref{fig:rho-T} its temperature variation is presented for several values of $J$ and in Fig.~\ref{fig:rho-J} $\rho$ is shown as a function of $J$ at a fixed temperature $T=0.2$, for three different values of the lattice size $L$. Vortices unbind close to the transition temperatures at which the magnetization rapidly declines (Fig.~\ref{fig:q1-T}) and the density of vortices rapidly increases. Examples of bound (unbound) vortex states inside (outside) the ferromagnetically ordered phase are shown in Fig.~\ref{fig:rho-J}. It is interesting to notice that $\rho$ is practically independent of the lattice size and it acquires the V-shape form as a function of $J$ at a fixed $T$. In particular, the vortex density decreases with the increasing competition between the nematic couplings and reaches a minimum close to $J=0.5$, i.e., $J_2=J_3$. 

\begin{figure}[t!]
\centering
\includegraphics[scale=0.5,clip]{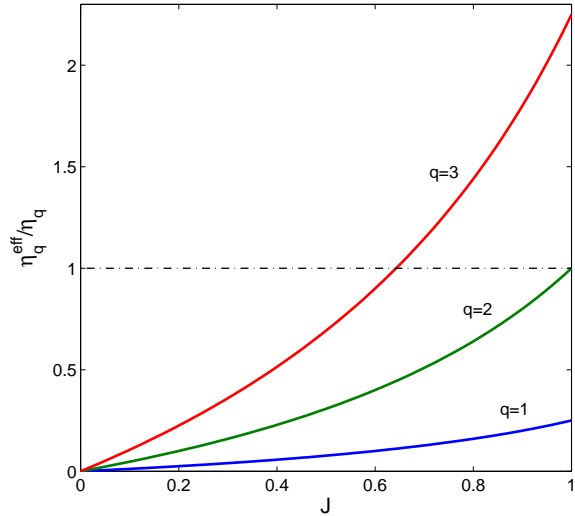}
\caption{(Color online) The reduced critical exponents $\eta_q^{eff}/\eta_q$ of the correlation functions $g_q$, $q=1,2,3$, versus $J$ in the SW approximation. The dashed line shows the value of the standard $XY$ model.}\label{fig:eta-J_sw}
\end{figure}

In order to quantify a joint effect of the presence of multiple couplings $J_q$, $q=1,\hdots,n$, on the decay rate of the algebraic correlation function $g_q$ within the BKT phase, we evaluate the critical exponent $\eta_q^{eff}$ by the spin-wave (SW) analysis. Furthermore, to compare it with the cases of pure couplings $J_q$ in Fig.~\ref{fig:eta-J_sw} we show the reduced quantity $\eta_q^{eff}/\eta_q = q^2J/(9-5J)$ as a function of $J$. We note that the presented results are valid for any temperature as the latter drops out from the ratio. Providing that the nature of the respective correlation functions remains algebraic for any value of $J$, one can observe that particularly the decay of the correlation function $g_1$ can be strongly suppressed ($\eta_q^{eff}/\eta_q \ll 1$). Nevertheless, we know that $g_1$ in the limiting cases of $J=0$ and $1$ does not decay algebraically but exponentially. Therefore, in the following we apply MC simulations to find out whether or not there is a region of $0 < J < 1$ of the slowly decaying $g_1$, as predicted by the SW approximation.

\begin{figure}[t]
\centering
    \subfigure{\includegraphics[scale=0.45]{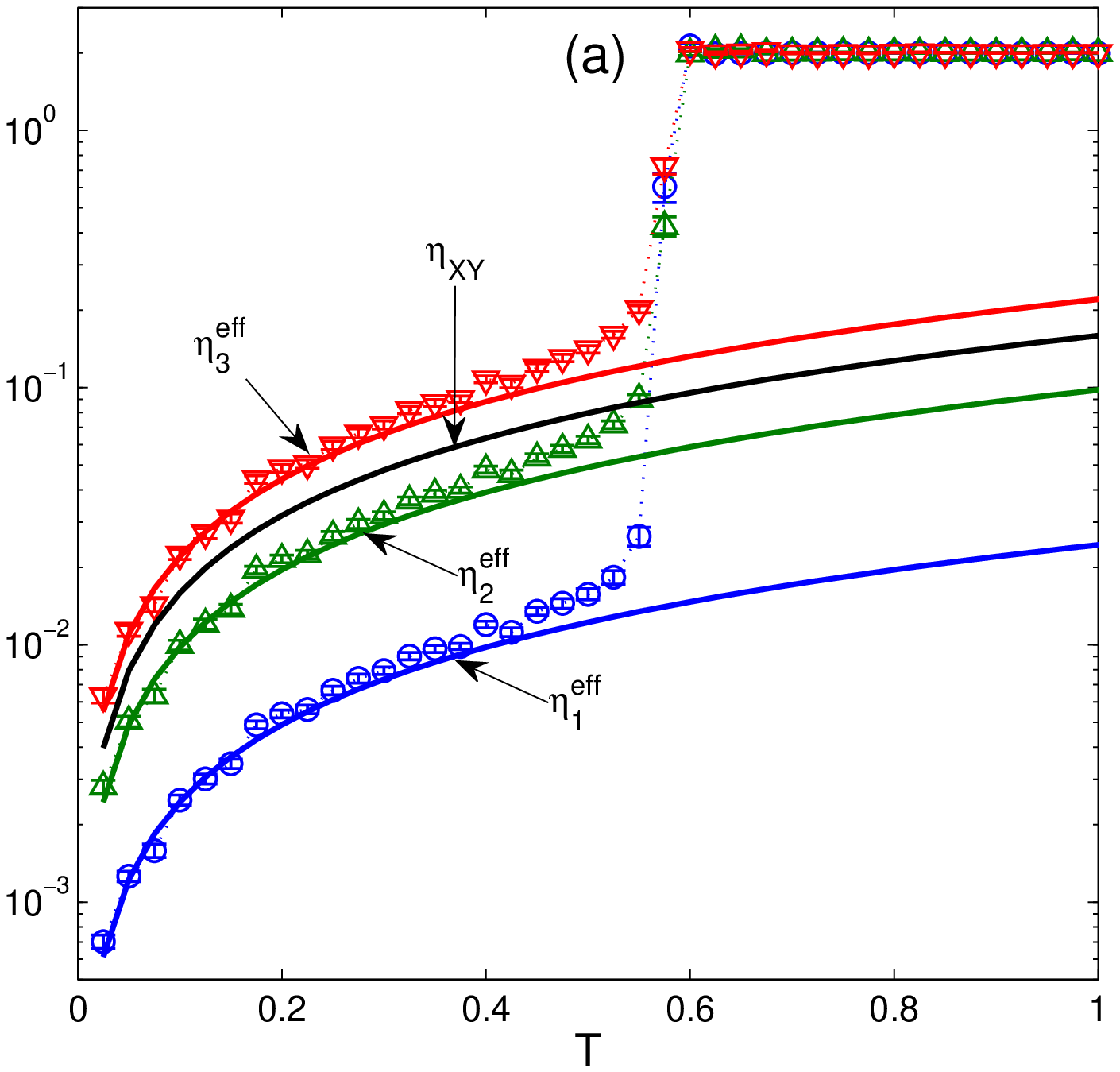}\label{fig:fss_all-T_J05}}
		\subfigure{\includegraphics[scale=0.45]{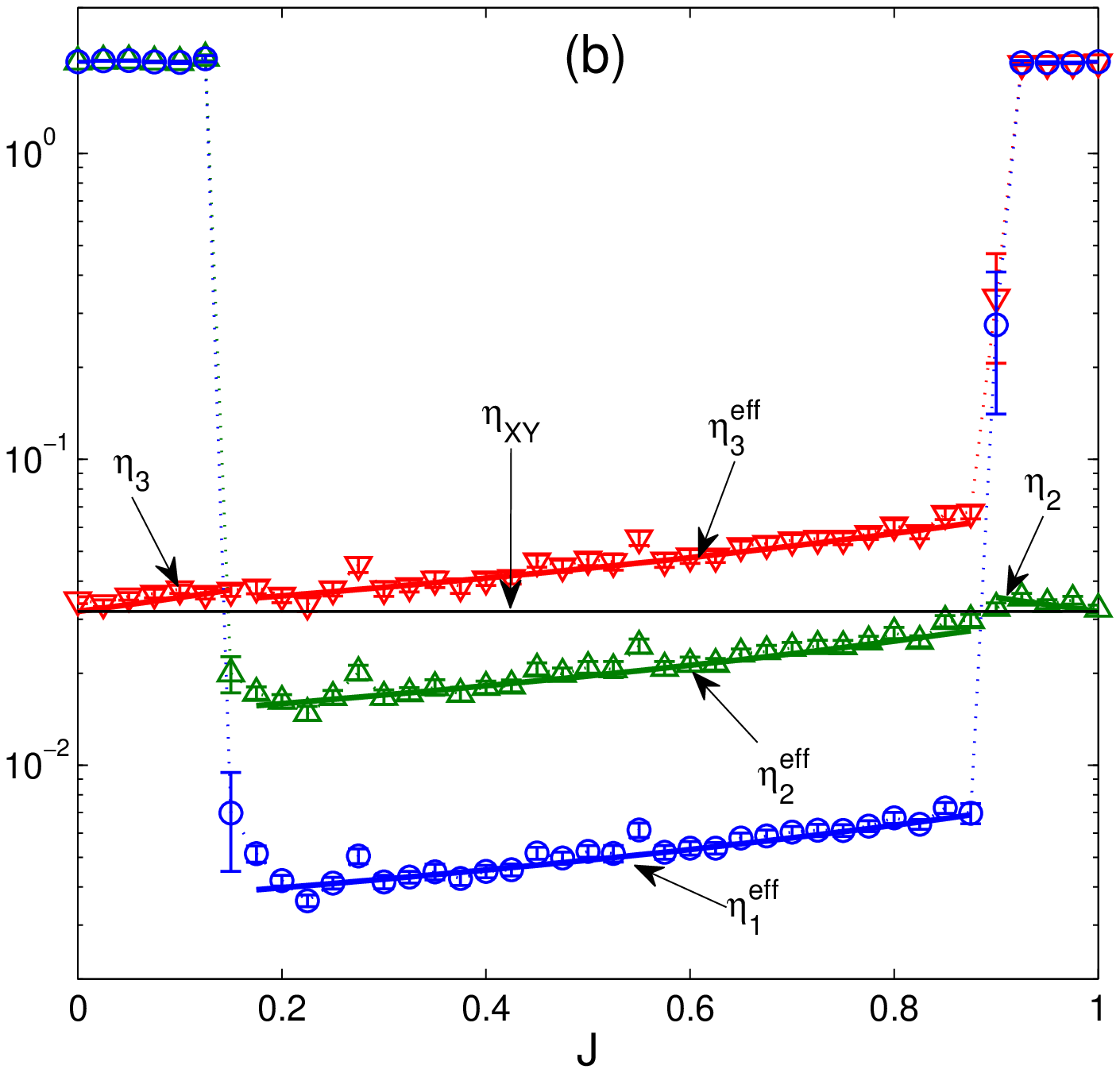}\label{fig:fss_all-J_T02}}
\caption{(Color online) The exponents $\eta_q^{eff}$, $q=1,2,3$, as a function of (a) temperature for $J=0.5$ and (b) the coupling constant $J$ for $T=0.2$, obtained from MC simulations (symbols) and the SW theory (solid lines). $\eta_{XY}$ denotes the SW approximation for the standard $XY$ model.}
\label{fig:fss_all}
\end{figure}

In Fig~\ref{fig:fss_all} we present the exponents $\eta_q^{eff}$, $q=1,2,3$, both as functions of temperature, for a fixed value of $J=0.5$, and as functions of the coupling constant $J$, for a fixed value of $T=0.2$. The symbols represent values obtained from MC simulations and the solid lines the predictions from the SW theory. First of all, the MC results clearly confirm the algebraic character of the decay of the respective correlation functions, including $g_1$, and also show that the SW theory gives good approximations in a substantial part of the low-temperature BKT phase. Then, by comparing the exponent $\eta_1^{eff}$ with $\eta_1 \equiv \eta_{XY}$ (black solid line), one can conclude that the magnitude of $\eta_1^{eff}$ is about one order smaller than $\eta_{XY}$. As $J$ approaches the limiting values of $0$ and $1$ the character of the decay of $g_1$ changes to the exponential within some intervals (see Fig.~\ref{fig:fss_all-J_T02}), the widths of which increase with $T$.  

\begin{figure}[t!]
\centering
\includegraphics[scale=0.52,clip]{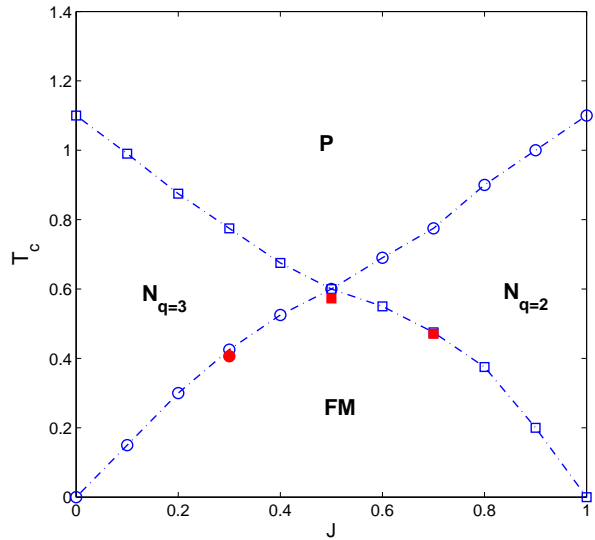}
\caption{(Color online) Phase diagram as a function of the parameter $J$. $FM$, $N_{q=2}$, $N_{q=3}$, and $P$ denote respectively the ferromagnetic, $q=2$ nematic, $q=3$ nematic and paramagnetic phases. The (pseudo)transition temperatures (empty symbols) are obtained from maxima of the specific heat curves, for $L = 24$, and the transition temperatures at $J=0.3,0.5$ and $0.7$ (filled red symbols) were determined from FSS.}\label{fig:PD}
\end{figure} 

The resulting phase diagram as a function of $J$ is presented in Fig.~\ref{fig:PD}. The boundaries marked by the empty symbols represent (pseudo)transition temperatures estimated from maxima of the specific heat curves, for the fixed value of $L = 24$. The transition temperatures at $J=0.3,0.5$ and $0.7$, marked by the filled red symbols, were determined more precisely from the FSS analysis and provide us an idea about the deviation between the (pseudo)transition and true transition temperatures. The phase boundaries split the parameter space into one disordered and three ordered phases. $P$ is the disordered paramagnetic phase and the ordered phases represent the QLRO states with the power-law decaying correlation functions $g_q,q=1,2,3$, $(FM)$, $g_2$ $(N_{q=2})$ and $g_3$ $(N_{q=3})$.

\begin{figure}[t!]
\centering
    \subfigure{\includegraphics[scale=0.42]{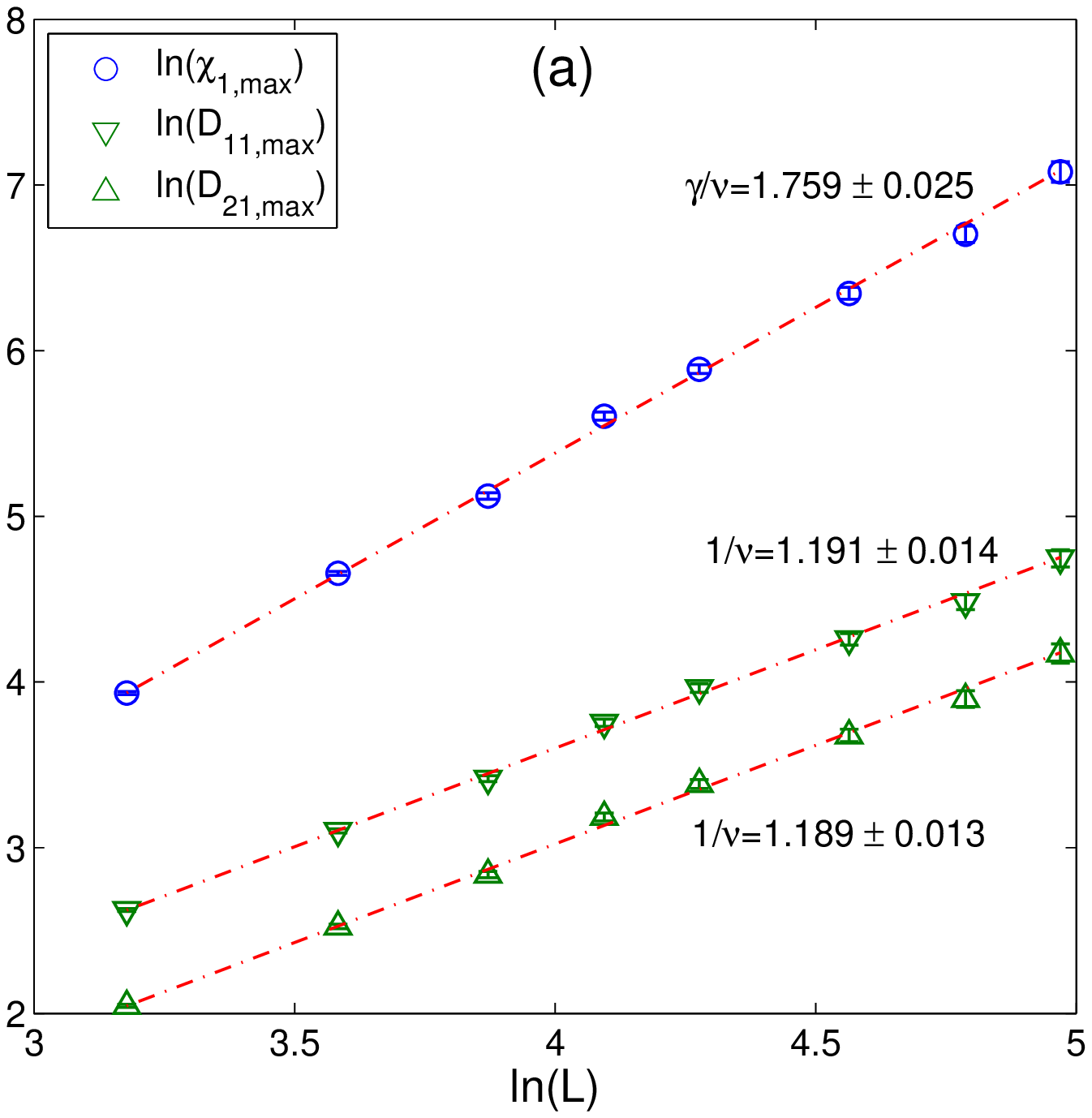}\label{fig:fss_T0_J_0_3_m_L144}}
	  \subfigure{\includegraphics[scale=0.42]{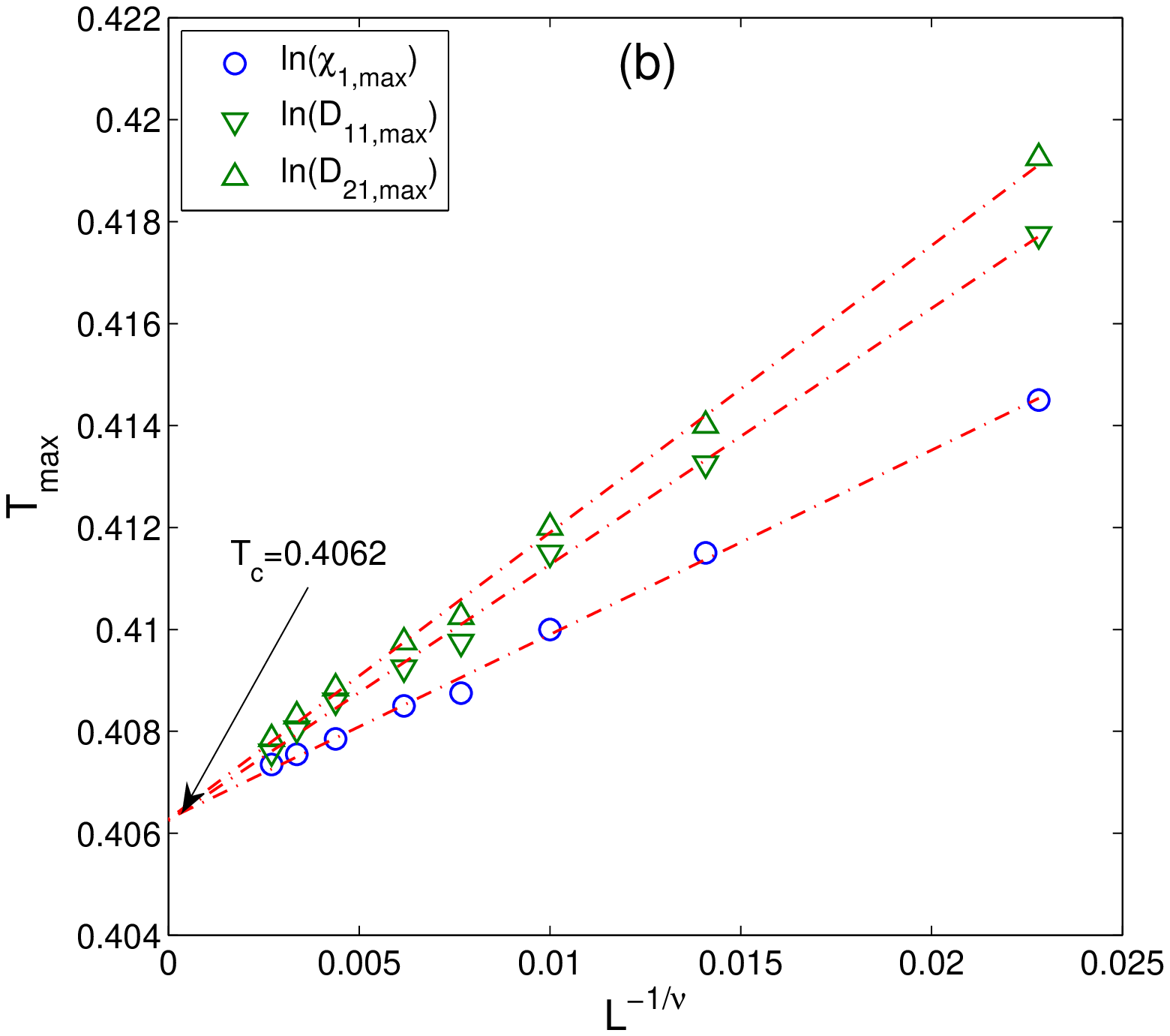}\label{fig:fss_Tc_J_0_3_m_L144}}
		\subfigure{\includegraphics[scale=0.42]{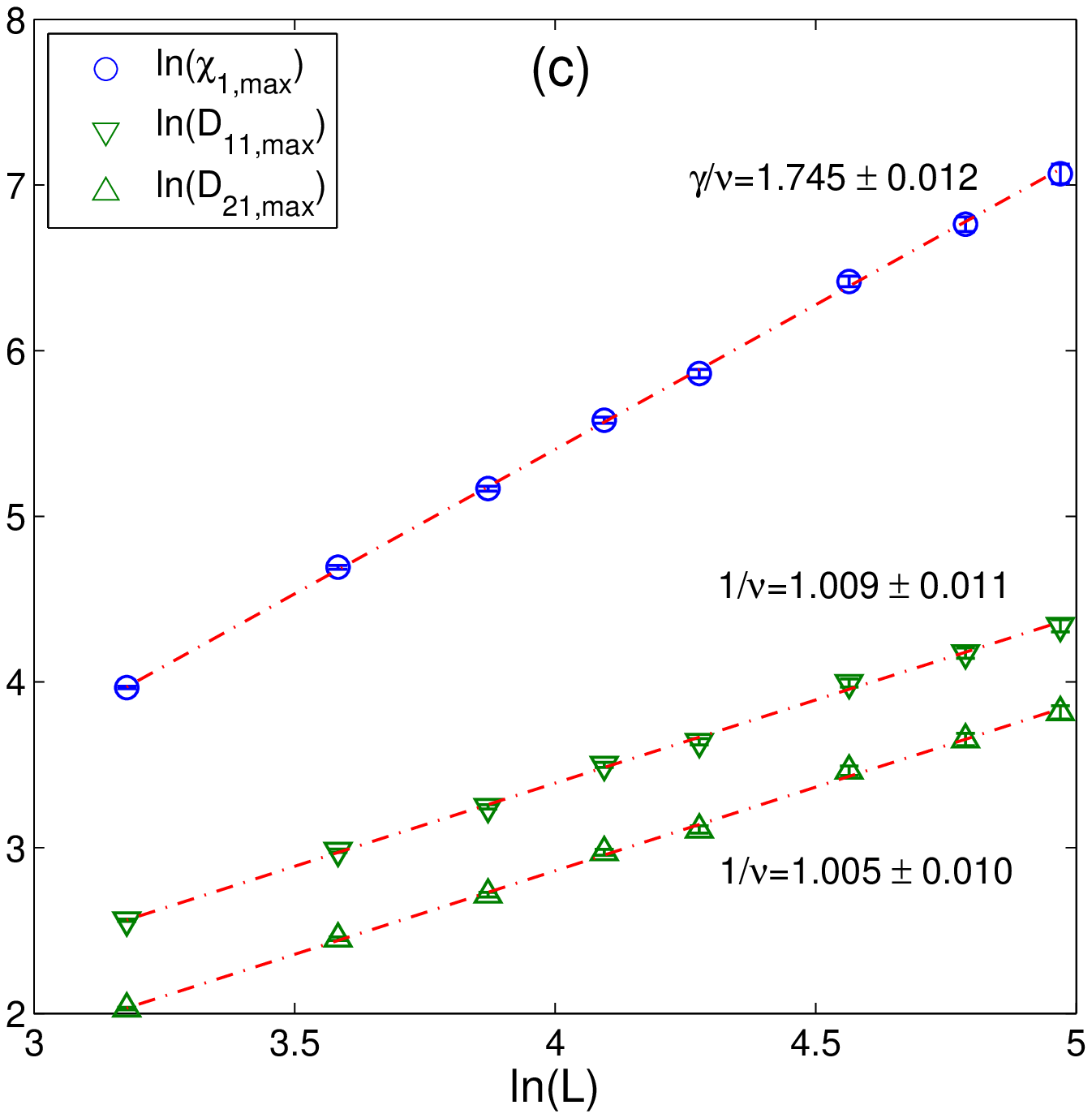}\label{fig:fss_T0_J_0_7_m_L144}}
		\subfigure{\includegraphics[scale=0.42]{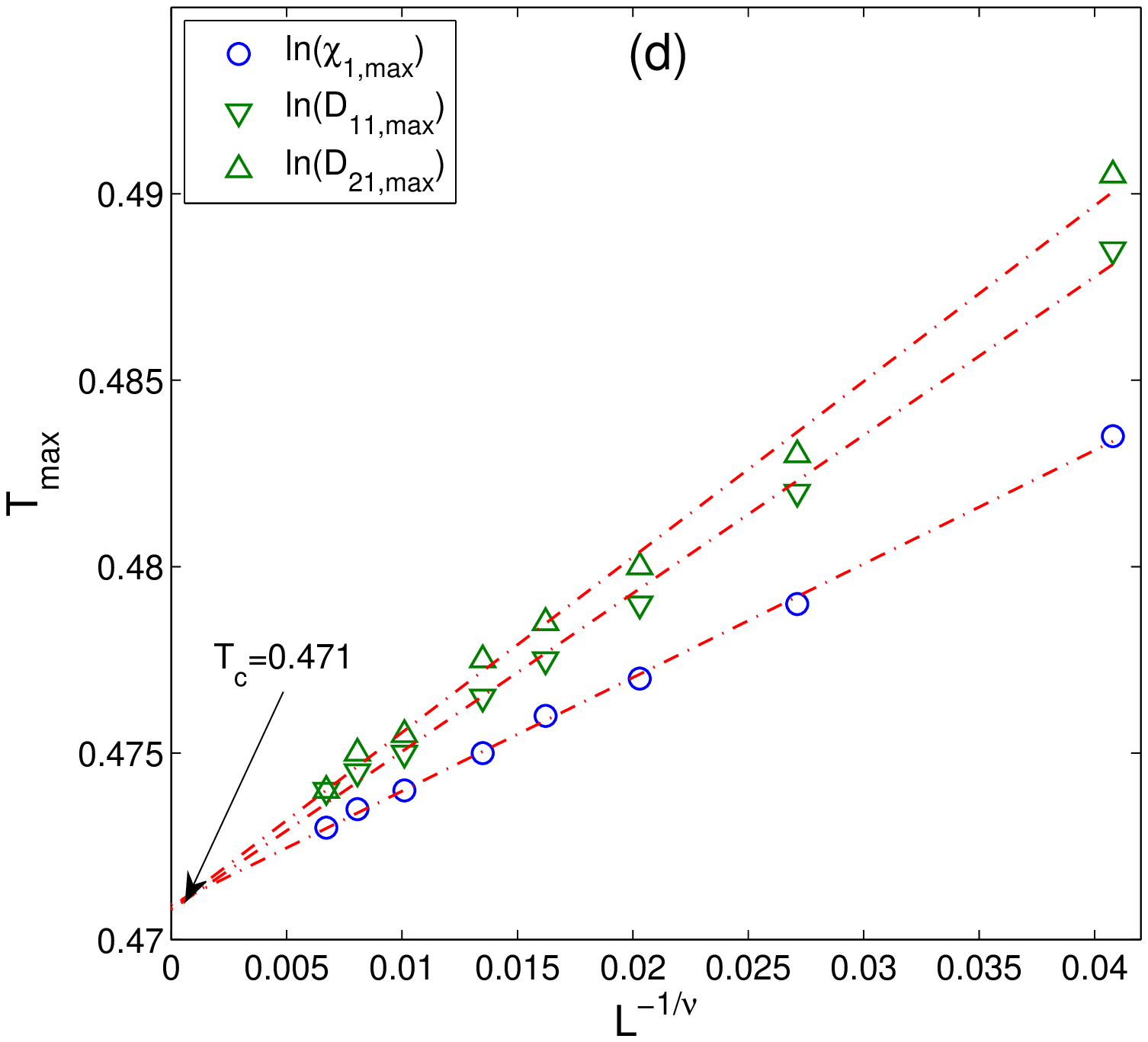}\label{fig:fss_Tc_J_0_7_m_L144}}
		\subfigure{\includegraphics[scale=0.42]{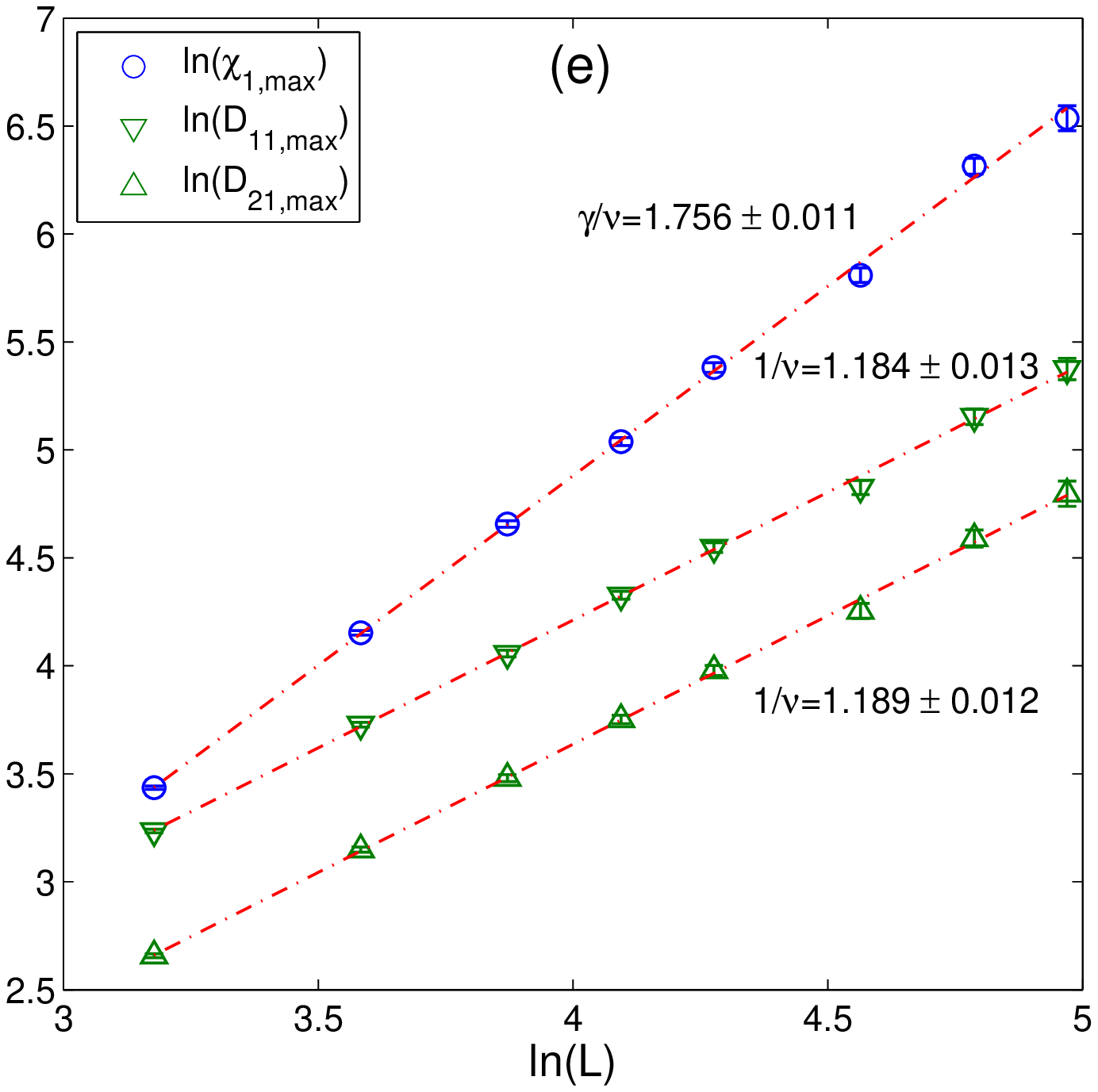}\label{fig:fss_T0_J_0_5_m_L144}}
		\subfigure{\includegraphics[scale=0.42]{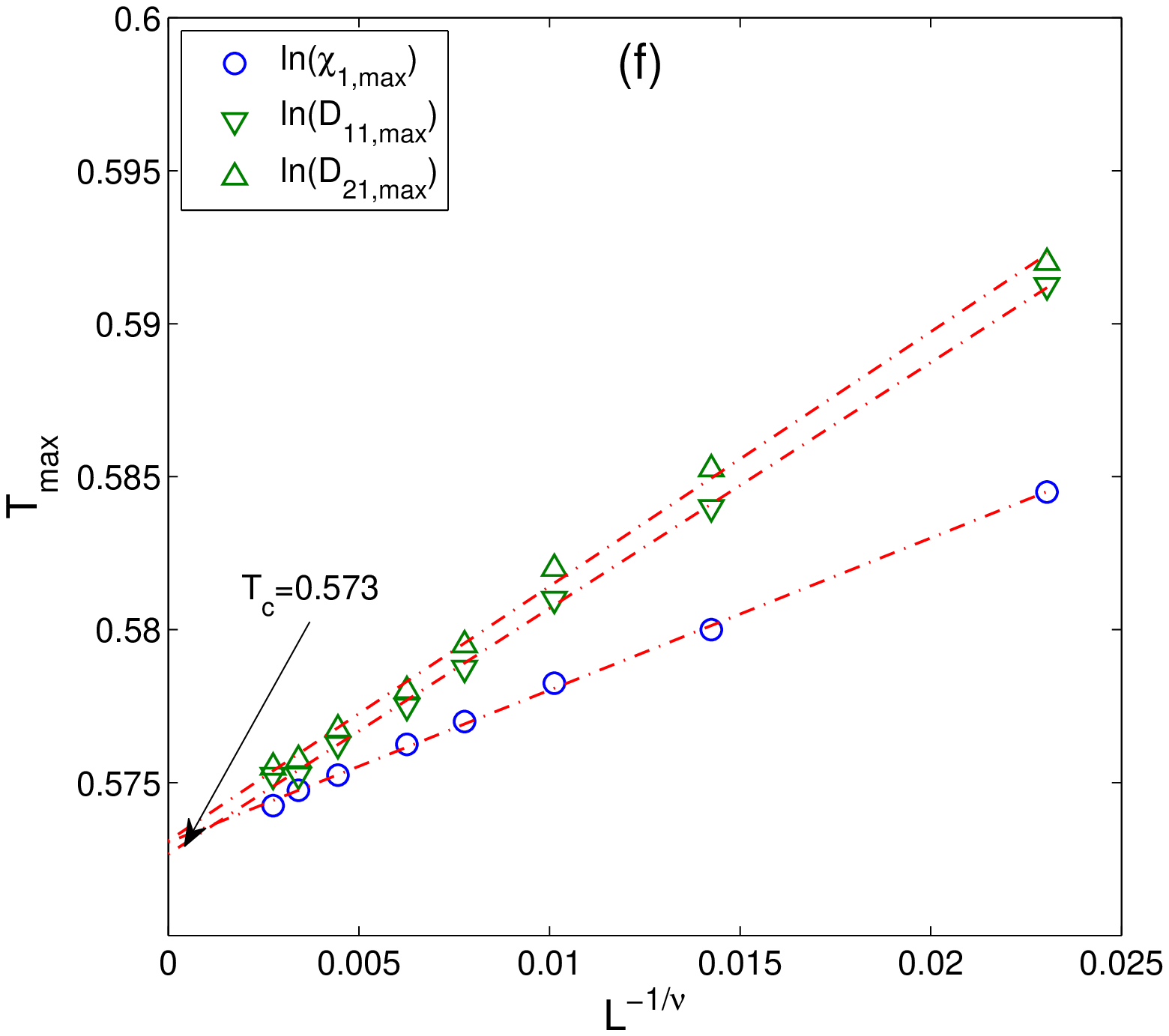}\label{fig:fss_Tc_J_0_5_m_L144}}
\caption{(Color online) Left (right) column: Critical exponents ratios (critical temperatures) at (a,b) the $FM-N_{q=3}$ ($J=0.3$), (c,d) $FM-N_{q=2}$ ($J=0.7$), and (e,f) $FM-P$ ($J=0.5$) phase transitions. The arrows in (b), (d) and (f) indicate the estimates of $T_c$ in the thermodynamic limit.}
\label{fig:fss}
\end{figure}

Finally, we focus on the nature of the phase transitions between the respective ordered phases. In particular, we apply the FSS analysis (Eqs.~\ref{fss_chi}-\ref{fss_D2}) to obtain the critical exponents and determine the universality class of the transition. The FSS analysis is performed for the selected values of $J=0.3$ and $0.7$, corresponding to the $FM-N_{q=3}$ and $FM-N_{q=2}$ transition, respectively, and for $J=0.5$ in the vicinity of the multicritical point at which the boundaries cross. The results are presented in Fig.~\ref{fig:fss}. For the $FM-N_{q=3}$ transition at $J=0.3$ the values of the critical exponents ratios $\gamma/\nu$ and $1/\nu$ point to the three-state Potts universality class with the values $\gamma_P/\nu_P=26/15 = 1.7{\bar 3}$ and $1/\nu_P=6/5=1.2$. On the other hand, for the $FM-N_{q=2}$ transition at $J=0.7$the obtained exponents ratios indicate the Ising universality class with the values $\gamma_I/\nu_I=7/4 = 1.75$ and $1/\nu_I=1$. These results are consistent with those obtained for the magnetic-nematic phase transition in the $J_1-J_2$ model~\cite{lee85,carp89,shi11,hubs13,qi13}, as well as for the the magnetic-nematic-like phase transition in the $J_1-J_3$ model~\cite{pode11,cano14}. The critical exponents at $J=0.5$ are found to be close to the three-state Potts universal values. The small deviations could be ascribed to the proximity of the muticritical point, which thus might be slightly different from $J=0.5$. Nevertheless, the estimated transition temperatures obtained for all the order parameters $o_q, q=1,2,3$ cannot be distinguished within the error bars (not shown).

\section{Summary and discussion}
We considered the $XY$ model involving solely nematic-like terms of the second (biquadratic) and third (bicubic) order, in the absence of the magnetic (bilinear) term, and studied the effect of their coexistence. By means of the spin-wave (SW) analysis and Monte Carlo (MC) simulation it was found that the mutual competition between them leads to a magnetic quasi-long-range ordering (QLRO). This phenomenon can only be attributed to the coexistence and competition between the nematic-like couplings as neither of them alone can induce such ordering. The resulting ferromagnetic phase ($FM$) is wedged between the nematic-like phases $N_{q=2}$ ($N_{q=3}$) with only axial spin alignments with the angles $0,\pi$ ($0,2\pi/3,4\pi/3$), in the limit of a relatively large biquadratic (bicubic) interaction. Thus, except the muticritical point, at which all the phases meet, for any value of the coupling parameter $0 < J< 1 $ there are two phase transitions: first from the paramagnetic phase to one of the two nematic-like phases followed by the second one at lower temperatures to the $FM$ phase.  

By applying a finite-size scaling (FSS) analysis of MC data on the boundaries between the magnetic and nematic-like phases we obtained for selected parameter values the critical exponents. The latter indicated that the phase transitions between the magnetic and nematic-like phases belong to the Ising and three-state Potts universality classes for the $FM-N_{q=2}$ and $FM-N_{q=3}$ transitions, respectively.  The FSS analysis performed inside the competition-induced QLRO magnetic phase, supported by the SW predictions, revealed that the spin-pair correlation function $g_1(r) \sim r^{-\eta_{1}^{eff}}$ decays even much more slowly than in the standard $XY$ model with a purely magnetic interaction, i.e., $\eta_{1}^{eff} \ll \eta_{XY}$. Such a magnetic phase is characterized by an extremely low vortex-antivortex pair density attaining a minimum close to $J=0.5$, i.e., the point at which the biquadratic $J_2$ and bicubic $J_3$ couplings are of about equal strengths and thus competition is the fiercest.

\begin{figure}[t!]
\centering
\includegraphics[scale=0.52,clip]{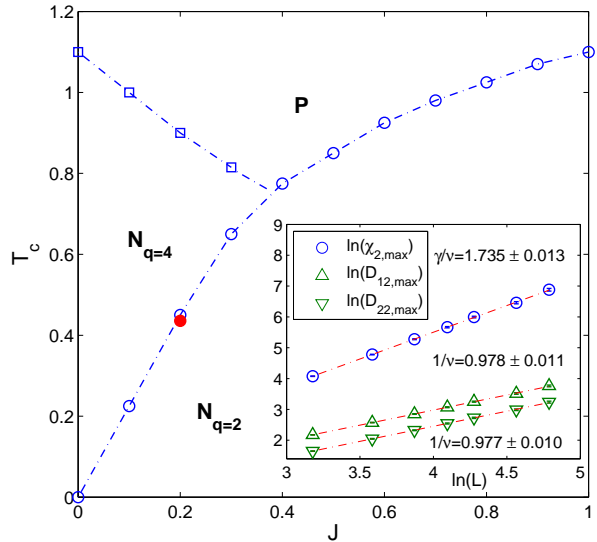}
\caption{(Color online) Phase diagram of the model ${\mathcal H}={\mathcal H_{2}}+{\mathcal H_{4}}$ as a function of the parameter $J$, where $J_2 \equiv J$ and $J_4=1-J$. $N_{q=2}$, $N_{q=4}$, and $P$ denote respectively the $q=2$ nematic, $q=4$ nematic and paramagnetic phases. The inset shows the FSS results performed at $J=0.2$ with the estimated critical temperature marked by the filled red circle.}\label{fig:PDq24}
\end{figure} 

We believe that nematic models including different higher-order couplings, with the generalized Hamiltonian ${\mathcal H}={\mathcal H_{q_1}}+{\mathcal H_{q_2}}$, where $q_1,q_2 \geq 2$, can bring about new competition-driven phases, similar to the model ${\mathcal H}={\mathcal H_1}+{\mathcal H_q}$, where $q \geq 5$~\cite{pode11,cano14,cano16}. Different higher-order couplings may compete, like in the present case, but may also collaborate. The latter case includes the model with $q_1=2$ and $q_2=4$. Both interactions enforce different but noncompeting noncollinear spin orientations $0,\pi$ and $0,\pi/2,\pi,3\pi/2$. Our preliminary calculations indicate that the topology of the resulting phase diagram is the same as in the bilinear-biquadratic model but the magnetic and nematic phases are replaced by the nematic $N_{q=2}$ and $N_{q=4}$ phases, respectively, and the transition between them belongs to the Ising universality class (see Fig.~\ref{fig:PDq24}).

\begin{acknowledgments}
This work was supported by the Scientific Grant Agency of Ministry of Education of Slovak Republic (Grant No. 1/0331/15) and the scientific grants of Slovak Research
and Development Agency provided under contracts No.~APVV-16-0186 and No.~APVV-14-0073.
\end{acknowledgments}


\begin{thebibliography}{30}
\bibitem{bere71} V. L. Berezinskii, Sov. Phys. JETP {\bf 34}, 610 (1972).
\bibitem{kost73} J. M. Kosterlitz and D. J. Thouless, J. Phys. C {\bf 6}, 1181 (1973); J. M. Kosterlitz, ibid. {\bf 7}, 1046 (1974).
\bibitem{carme87} H.-O. Carmesin, Phys. Lett. A {\bf 125}, 294 (1987).
\bibitem{lee85} D. H. Lee and G. Grinstein, Phys. Rev. Lett. {\bf 55}, 541 (1985).
\bibitem{kors85} S. E. Korshunov, Pis’ma Zh. Eksp. Teor. Fiz. 41, 216 (1985) [JETP Lett. {\bf 41}, 263 (1985)].
\bibitem{geng09} J. Geng and J. V. Selinger, Phys. Rev. E {\bf 80}, 011707 (2009)
\bibitem{hlub08} R. Hlubina, Phys. Rev. B {\bf 77}, 094503 (2008).
\bibitem{grason08} G. M. Grason, Europhysics Letters {\bf 83}, 58003 (2008).
\bibitem{bonnes12} L. Bonnes and S. Wessel, Phys. Rev. B 85, 094513 (2012).
\bibitem{bhas12} M. J. Bhaseen, S. Ejima, F. H. L. Essler, H. Fehske, M.
Hohenadler, and B. D. Simons, Phys. Rev. A {\bf 85}, 033636 (2012).
\bibitem{forg16} L. de Forges de Parny, A. Rancon, and T. Roscilde, Phys. Rev.
A {\bf 93}, 023639 (2016).
\bibitem{cairns16} A. B. Cairns,	M. J. Cliffe, J. A. M. Paddison,	D. Daisenberger, M. G. Tucker,	F.-X. Coudert,	and A. L. Goodwin, Nature Chemistry {\bf 8}, 442 (2016).
\bibitem{zuko16} M. \v{Z}ukovi\v{c}, Phys. Rev. B {\bf 94}, 014438 (2016).
\bibitem{carp89} D. B. Carpenter and J. T. Chalker, Journal of Physics: Condensed Matter {\bf 1}, 4907 (1989).
\bibitem{shi11} Y. Shi, A. Lamacraft, and P. Fendley, Phys. Rev. Lett. {\bf 107}, 240601 (2011).
\bibitem{hubs13} D. M. H\"{u}bscher and S. Wessel, Phys. Rev. E {\bf 87}, 062112 (2013).
\bibitem{qi13} K. Qi, M. H. Qin, X. T. Jia, and J.-M. Liu, J. Magn. Magn. Mater. {\bf 340}, 127–130 (2013).
\bibitem{pode11} F. C. Poderoso, J. J. Arenzon, and Y. Levin, Phys. Rev. Lett. {\bf 106}, 067202 (2011).
\bibitem{cano14} G. A. Canova, Y. Levin, and J. J. Arenzon, Phys. Rev. E {\bf 89}, 012126 (2014).
\bibitem{cano16} G. A. Canova, Y. Levin, and J. J. Arenzon, Phys. Rev. E {\bf 94}, 032140 (2016).
\bibitem{fari05} A. I. Fari\~{n}as-S\'{a}nchez, R. Paredes, and B. Berche, Phys. Rev. E {\bf 72}, 031711 (2005).
\bibitem{berc05} B. Berche and R. Paredes, Condensed Matter Physics {\bf 8}, 723–736 (2005).
\bibitem{doma84} E. Domany, M. Schick, and R. H. Swendsen, Phys. Rev. Lett. {\bf 52}, 1535 (1984).
\bibitem{himb84} J. E. Van Himbergen, Phys. Rev. Lett. {\bf 53}, 5 (1984).
\bibitem{blot02} H. W. J. Bl\"{o}te, W. Guo, and H. J. Hilhorst, Phys. Rev. Lett. {\bf 88}, 047203 (2002).
\bibitem{sinh10a} S. Sinha and S. K. Roy, Phys. Rev. E {\bf 81}, 022102 (2010).
\bibitem{sinh10b} S. Sinha and S. K. Roy, Phys. Rev. E {\bf 81}, 041120 (2010).
\bibitem{zuko17} M. \v{Z}ukovi\v{c} and G. Kalagov, Phys. Rev. E {\bf 96}, 022158 (2017).
\bibitem{ente02} A. C. D. van Enter and S. B. Shlosman, Phys. Rev. Lett. {\bf 89}, 285702 (2002).
\bibitem{ente05} A. C. D. van Enter and S. B. Shlosman, Comm. Math. Phys. {\bf 255}, 21 (2005).
\bibitem{ferr88} A. M. Ferrenberg and R. H. Swendsen, Phys. Rev. Lett. {\bf 61}, 2635 (1988).
\bibitem{ferr89} A. M. Ferrenberg and R. H. Swendsen, Phys. Rev. Lett. {\bf 63}, 1195 (1989).
\bibitem{wolf04} U. Wolff, Computer Physics Communications {\bf 156}, 143  (2004).
\bibitem{fish73} M. E. Fisher, M. N. Barber, and D. Jasnow, Phys. Rev. A {\bf 8}, 1111 (1973).
\bibitem{nels77} D. R. Nelson and J. M. Kosterlitz, Phys. Rev. Lett. {\bf 39}, 1201 (1977).
\bibitem{minn03} P. Minnhagen and B. J. Kim, Phys. Rev. B {\bf 67}, 172509 (2003).
\bibitem{jin12} S. Jin, A. Sen, and A. W. Sandvik, Phys. Rev. Lett. {\bf 108}, 045702 (2012).

\end{thebibliography}
\end{document}